# Physically informed machine-learning algorithms for the identification of two-dimensional atomic crystals

Laura Zichi[1], Tianci Liu[2], Elizabeth Drueke[1], Liuyan Zhao[1,*], and Gongjun Xu[2,*]

[1]University of Michigan, Department of Physics, Ann Arbor, 48109, United States of America
[2]University of Michigan, Department of Statistics, Ann Arbor, 48109, United States of America

*lyzhao@umich.edu
*gongjun@umich.edu

## Abstract

First isolated in 2004, graphene monolayers display unique properties and promising technological potential in next generation electronics, optoelectronics, and energy storage. The simple yet effective methodology, mechanical exfoliation followed by optical microscopy inspection, used for fabricating graphene has been exploited to discover many more two-dimensional (2D) atomic crystals which show distinct physical properties from their bulk counterpart, opening the new era of materials research. However, manual inspection of optical images to identify 2D flakes has the clear drawback of low-throughput and hence is impractical for any scale-up applications of 2D samples, albert their fascinating physical properties. Recent integration of high-performance machine-learning, usually deep learning, techniques with optical microscopy has accelerated flake identification. Despite the advancement brought by deep learning algorithms, their high computational complexities, large dataset requirements, and more importantly, opaque decision-making processes limit their accessibilities. As an alternative, we investigate more transparent tree-based machine-learning algorithms with features that mimic color contrast for the automated identification of exfoliated 2D atomic crystals (e.g., $MoSe_2$) under different optical settings. We compare the success and physical nature of the decisions of these tree-based algorithms to ResNet, a Convolutional Neural Network (CNN). We show that decision trees, gradient boosted decision trees, and random forests can successfully classify optical images of thin materials with transparent decisions that rely on physical image features and do not suffer from extreme overfitting and large dataset requirements.

## Introduction

Since the first realization in 2004 that graphite could be mechanically exfoliated into graphene in ambient conditions using a simple piece of Scotch tape[1], the study of graphene in the two-dimensional (2D) limit has demonstrated a rich landscape for interesting physical phenomena, ranging from Dirac electrons in graphene[2] to unconventional superconductivity in twisted graphene moire superlattices[3]. This simple yet effective methodology used for fabricating graphene, scotch tape peeling followed by optical microscope imaging, has been exploited to greatly expanded the 2D atomic crystal pool, discovering 2D transition metal dichalcogenide (TMD) semiconductors[4] 2D magnets[5] and many on. However, many of these materials show deterioration in ambient conditions[6-8]. This has prompted researchers to innovate fabrication environments which prevent degradation of samples. Further improvements to current fabrication and visualization techniques to produce large 2D materials remains imperative for their fundamental research and commercial-level applications in next generation electronics, optoelectronics, and energy storage[9,10].

Isolation of 2D materials involves cleaving the bulk material on a wafer, usually oxidized Si. Current visualization techniques, including atomic-force, scanning tunneling, and electron microscopies, exhibit low-throughput of locating the resultant relevant thin materials of only several nm in diameter known as flakes[11]. Raman Microscopy, which can realize accurate 2D structures, has not been automated and relies on experienced users[12]. With the recent surge of successful high-performance machine-learning algorithms for object classification within images, many have applied these methods to locate exfoliated 2D materials in optical images[13-16]. Usually the machine-learning algorithms used for flake identification are deep neural networks due to their great success with object recognition[17-19]. The integration of machine-learning and optical microscopy techniques can accelerate flake identification. This can then expedite innovations within the flake fabrication process to promote practical 2D material applications and research.

However, neural networks' high computational complexity and large dataset requirements compared to more traditional tree-based algorithms render them unattainable and unsuitable for certain environments. Furthermore, critiqued as "black boxes", no comprehensive theoretical understanding of neural networks' inner layers exists[20]. Therefore, we propose coupling optical microscopy techniques with tree-based algorithms as an alternative to deep learning methods for a more accessible and transparent automated identification of 2D materials. We employed, for comparison, tree-based methods - decision trees, gradient boosted decision trees and random forests - and deep Convolutional Neural Networks (CNNs) for identification of exfoliated $MoSe_2$ under different optical settings. The tree-based algorithm's features mimicked the physical method of identifying flakes using color contrast, a technique currently used throughout the 2D materials community, giving them a more understandable physical motivation than the CNNs[11]. We compare the physicality of these algorithms through tree visualizations and Gradient-weighted Class Activation Mapping (Grad-CAM), a post-hoc study that identifies the sub-regions CNNs rely on for classification, and their accuracies to understand their potential application[21,22].

## Methods

Optical images used in this study include transition metal dichalcogenide (TMD) flakes on $SiO_2$/Si substrate, TMD flakes on Polydimethylsiloxane (PDMS), and TMD flakes on $SiO_2$/Si and PDMS (if any). The usage of multiple types of substrates models more realistic flake fabrication environments and strengthens algorithm robustness. All these samples were mechanically exfoliated in a 99.999% $N_2$-filled glove box (Fig. 1 a.). The optical images were also acquired in the same environment with no exposure to ambient conditions occurring between fabrication and imaging processes (Fig. 1 b.). The 83 $MoSe_2$ images used throughout this study were taken at the 100× magnification setting with varied microscope settings, selected contrast, etc (Fig. 1 c.). These images are divided into four smaller symmetric images containing randomized amounts of flake and bulk material which were then manually reclassified (Fig. 1 d.).

The extremely time-consuming process of locating a flake renders these datasets small, a common occurrence in many domains such as medical sciences and physics. However, deep learning models, such as CNNs, usually contain numerous parameters to learn and require large-scale data to train on to avoid severe overfitting. Data augmentation is a practical solution to this problem[23]. By generating new samples based on existing data, data augmentation produces training data with boosted diversity and sample sizes, on which better performing deep learning models can be trained (see Supplementary methods). The benefit of applying data augmentation is two-fold. First, it enlarges the data that CNNs are trained on. Second, the randomness induced by the augmentation of the data encourages the CNNs to capture and extract spatially invariant features to make predictions, improving the robustness of the models[23]. In fact, augmentation is quite common when using CNNs even with large datasets for this reason. Typically, different augmented images are generated on the fly during the model training period, which further helps models to extract robust features. Due to limited computing resources, we generated augmented data prior to fitting any models, expanding the data from 332 to 10,292 images (Fig. 1 e.).

Once augmented, we applied color quantization to all images (Fig. 1 f.). The quantization decreased noise and image colors to a manageable number necessary for extracting the tree-based algorithms' features. The color quantization algorithm uses a pixel-wise Vector Quantization to reduce colors within the image to a desired quantity while preserving the original quality[16]. We employed a K-means clustering to locate the desired number of color cluster centers using a single byte and pixel representation in 3D space. The K-means clustering trains on a small sample of the image and then predicts the color indices for the rest of the image, recreating it with the specified number of colors (see Supplementary methods). We recreated the original $MoSe_2$ images with 5, 20, and 256 colors to examine which resolution produced the most effective and generalizable models. The lower edge of five represented the lowest number of colors that would not regularly remove small flakes from recreated images. Images recreated with 20 colors appeared almost indistinguishable from the original while still greatly decreasing noise. To mimic an unquantized image, we recreated images with 256 color clusters. We compare the accuracies of the tree-based algorithms and CNNs on datasets of our images recreated with 5 and 20 colors. We also compare the tree-based algorithms performance on our images recreated with 256 colors to the CNNs on the unquantized images (it is not necessary to perform quantization for CNN classification).

After processing the optical images, we employ tree-based and deep learning algorithms for their classification. Tree-based algorithms are a family of supervised machine learning that perform classification or regression based on the value of the features of the tree-like structure it constructs. A tree consists of an initial root node, decision nodes which produce binary answers based on the tree's features, and childless leaf nodes (or terminal nodes) where a target variable class or value is assigned [24]. Decision trees' various advantages include the ability to successfully model complex interactions with discreet and continuous attributes, high generalizability, robustness to predictor variable outliers, and an easily interpreted decision-making process[25,26]. These attributes motivate the coupling of tree-based algorithms and optical microscopy for automated identification of 2D materials. Specifically, we employ decision trees along with ensemble classifiers, such as random forests and gradient boosted decision trees, for improved prediction accuracies and smoother classification boundaries[27-29].

The features of the single and ensemble trees mimic the physical method of using color contrast for identifying graphene crystallites against a thick background. The flakes are sufficiently thin so that their interference color will differ from an empty wafer, creating a visible optical contrast for identification[11]. Calculated color differences, based on RGB color codes, between every combination of color clusters model optical contrast. These differences are sorted into bins which encompass data extrema and have finer ranges around a manually determined flake to background contrast range. To prevent model overfitting, especially for the ensemble classifiers, only three relevant color contrast ranges were chosen for training and test of the models: the lowest range, a middle range representative of the color contrast between a flake and background material, and the highest range (see Supplementary methods). Once these features are calculated, we employed a k-fold cross-validation grid search to determine the best values for each estimator's hyperparameters. The k-fold cross-validation – an iterative process that divides the train data into k partitions – uses one partition for validation (testing) and the remaining k-1 for training during each iteration[30]. For each tree-based method, the estimator with the combination of hyperparameters which produces the highest accuracy on the test data was selected (see Supplementary methods). We employed a 5-fold cross-validation with a standard 75/25 train/test split. After finetuning the decision tree's hyperparameters with k-fold cross-validation, we produced visualizations of the estimator to evaluate the physical nature of its decisions. The gradient boosted decision tree and random forest estimators represent ensembles of decision trees so the overall nature of their decisions can be extrapolated from a visualization of a single decision tree since they all use the same inherently physical features.

Along with the tree-based methods, we also examined deep learning algorithms. Recently, deep neural networks, which learn more flexible latent representations with successive layers of abstraction, have shown great success on a variety of tasks including object recognition[31,32]. Deep convolutional neural networks take an image as input and output a class label or other types of results depending on the goal of the task. During the feed forward step, a sequence of convolution and pooling operations are applied to the image to extract visual. The CNN model we employ is a ResNet18[33], and we train new networks from scratch by initializing parameters with uniform random variables[34] due to the lack of public neural networks pre-trained on similar data. The training of ResNet18 is as follows. We used 75% original images and all their augmented images as the training. This can further be split into training and validation sets when tuning hyper-parameters. We used a small batch size of 4 and run 50 epochs using stochastic gradient descent method with momentum[35]. We used a learning rate of 0.01 and momentum factor of 0.9. Various efforts work to produce accurate visualizations of the inner layers of CNNs including Grad-CAM which we employed. Grad-CAM does not give a complete visualization of the CNNs as it only uses information from the last convolutional layer of the CNN. However, this last convolutional layer is expected to have the best trade-off between high-level semantics and spatial information rendering Grad-CAMs successful in visualizing what CNNs use for decisions [21].

## Results and Discussions

Both machine-learning methods proved effective for an initial, automated identification of thin materials. The accuracies of the tree-based methods are displayed as both the average test score from the 5-fold cross-validation (blue) as well as the accuracy on the test dataset (green) as a function of the number of quantized colors (Fig. 2). The CNNs demonstrated higher accuracies than the tree-based methods for every color quantization. They performed with accuracies between 70.0-76.0% and showed no discernable dependency on color quantization. Conversely, the test and average 5-fold cross-validation accuracy of the tree-based methods improved as the images were quantized with more colors for all three algorithms. Unsurprisingly, the ensemble estimators, gradient boosted decision trees and random forests, performed with the highest accuracies of the tree-based methods. The accuracy of these methods ranged from 64.5-69.5% (Fig. 2).

Although the CNNs achieved higher accuracies, visualizations of the subregions they used for classification, through Grad-CAM images, indicate that the decision processes may lack physical integrity. Intuitively, the Grad-CAM uses gradients of the label *flake* to features (pixels) to locate and visualize the image subregions that the CNN used for training and testing during classification[21]. The final convolutional layer is used to construct a coarse heatmap indicating these subregions which is then overlaid onto the original image. We evaluated the Grad-CAM's ability to locate the flakes in 500 correctly classified images quantized with 20 colors with flakes to ascertain the physical nature of the CNNs. A masking algorithm located the region of each image containing a flake (see Supplementary methods). We then summed the Grad-CAM heatmap's weights (which are normalized between zero and one) in this region and divided by the total flake area. We summarized the results of this evaluation as an empirical cumulative distribution function (ECDF) shown in Fig. 3. We also showcase in Fig. 3 an example of a successful Grad-CAM image with an overlap fraction of 0.95 along with an unsuccessful

Grad-CAM image with 0.00 fractional overlap. The ECDF's median of 0.4 fractional overlap between a flake and a Grad-CAM image indicates that the CNN's did not use regions near the flake for training and test. Instead, the CNN's regularly failed to locate the flake, training and testing on other potentially meaningless image features while still correctly classifying images (Fig. 3 c.). This along with training accuracies that range from 99-100% indicates that the CNNs severely overfit on the training data which leads to poor generalization. Examining the performance of the CNNs on smaller training sets further underlines this. When trained with 10% and 50% of all training data from a 75/25 train test split, the CNNs suffered a large loss of accuracy (about 20%) while still maintaining training accuracies that range from 96-100%. Conversely, the tree-based methods maintained their performance within about 6% with training accuracies between 61-87% emphasizing their accessibility when working with limited data and diverse images (see Supplementary methods). The CNN's fortuitous ability to locate flakes and extreme overfitting emphasizes the need for caution when blindly applying these high-performance deep learning algorithms.

Conversely, the tree-based algorithms inherently relied on physical image attributes because their features are based on color contrast. Visualizations of the tree-based methods' decisions demonstrate their reliance on the physical color contrast of flake to bulk or background material for classification. To better highlight these ideas, we showcase how an image with a flake and without a flake are classified based on a decision tree trained on data quantized with 256 color cluster centers in Fig. 4. Each image's associated features of number of color differences in various ranges are shown as a bar chart below the images. The image without a flake traverses left from the leaf node and is classified as not containing a flake by a terminal node (Fig. 4 b). Starting from the root node, the image with a flake traverses the tree to the right until being classified as containing a flake at a terminal node (Fig. 4 c). The trees-based methods used the number of large color differences between pixels to identify images with flakes surrounded by bulk material (Fig. 4 b.). Similarly, these methods regularly used high numbers of low color differences to correctly classify optical images that are almost all background with little $MoSe_2$ as not containing flakes (Fig. 4 c.). Furthermore, the training accuracies ranged from 61-73% indicating little to no overfitting. However, the high Gini impurities associated with the various tree decision nodes indicates a lack of confidence in classification, highlighted by the tree-based estimator's lower accuracies.

These lower accuracies result from high rates of false negatives as indicated by the confusion matrices (see Supplementary methods). A manual post image processing revealed that the false negatives usually contained very small flakes. Further evaluation of the tree-based methods with operating characteristic (ROC) curves indicated that by tuning the true positive and false positive rates the tree-based methods can increase throughput of manual flake identification by a factor of three (see Supplementary methods). Although tree-based methods require fine tuning of features to increase accuracies, they represent a promising physically informed and transparent alternative to deep-learning algorithms for coupling with optical microscopy for rapid identification of thin materials.

## Acknowledgements

We acknowledge the Hui Deng group for sharing the optical images of 2D flakes. L.Zhao acknowledges the support by NSF CAREER Grant No. DMR1749774 and the Alfred P. Sloan foundation.

## Author Contributions

L.Zichi, E.D., and L.Zhao conceived the idea and initiated the project. L.Zichi implemented the tree-based methods, created the masking algorithm, evaluated the Grad-CAM images, wrote the original manuscript, and designed the figures. T.L. implemented the CNNs, augmented the data, created the Grad-CAM images, and optimized the color quantization algorithm. E.D. instructed the tree-based algorithm features and revised the manuscript. G.X. instructed all machine-learning algorithm evaluation. All authors reviewed the manuscript.

## Data availability

Optical images used for machine-learning training and testing are available upon request. All codes discussed here (machine-learning methods, masking algorithm, and Grad-CAM evaluation) are available on GitHub at https://github.com/lzichi/Thin-Materials-ML.

## Competing interests

All authors declare no competing interests.

Figure legends

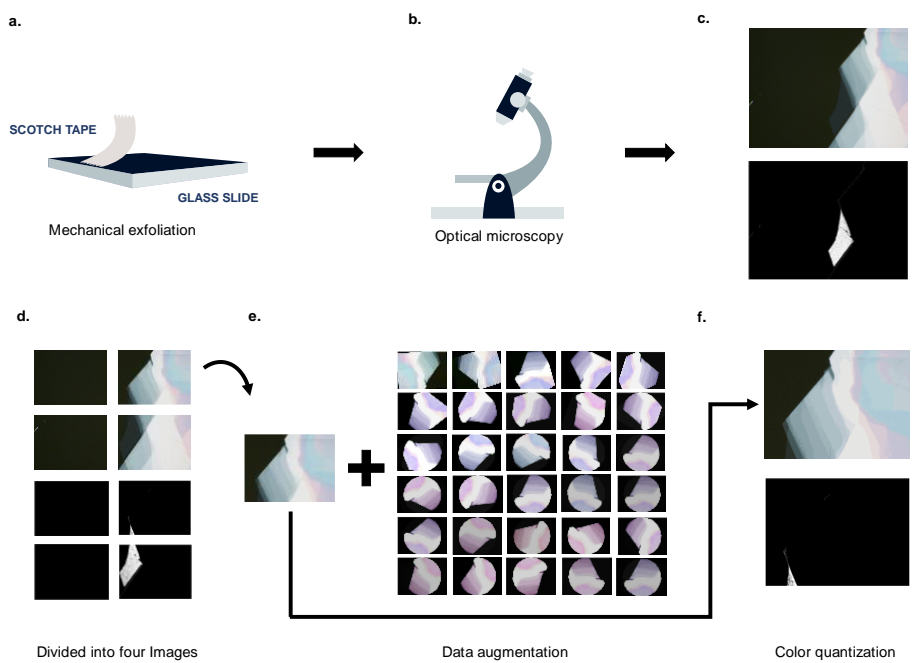

**Figure 1:** MoSe$_2$ flake fabrication and image collection and processing. (**a**) Mechanical exfoliation of MoSe$_2$ with scotch tape to produce flakes which are then (**b**) imaged with optical microscopy. (**c**) A typical optical image of a flake and surrounding bulk material with a masked version of the image below which only displays the flake in white. (**d**) The four resulting images when the original image in (**c**) is divided with the masked version below. (**e**) The resulting 30 images produced through the augmentation methods of padding, rotating, flipping, and color jitter. (**f**) The image recreated with 20 colors again with the masked version below.

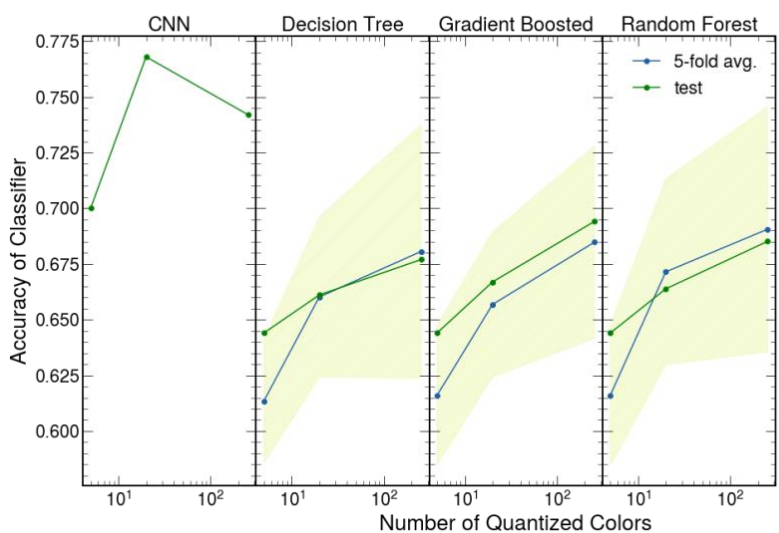

**Figure 2:** CNN and tree-based machine learning algorithms' accuracies. The accuracy of CNN and tree-based algorithms depends on the number of color clusters in the recreated image. The tree-based methods and CNNs were trained and tested on images recreated with 5, 20 and 256 colors and 5, 20, and infinite (original image) colors. The CNNs' accuracies (first panel) determined with a 75/25 train test split. Tree-based algorithms' accuracies (last three panels) shown as the average test accuracy from the 5-fold cross-validation used to select the algorithm's hyperparameters (blue) and test accuracy from 75/25 train test split after this optimization (green). The standard error for the 5-fold cross-validation test accuracy is represented by the shaded region.

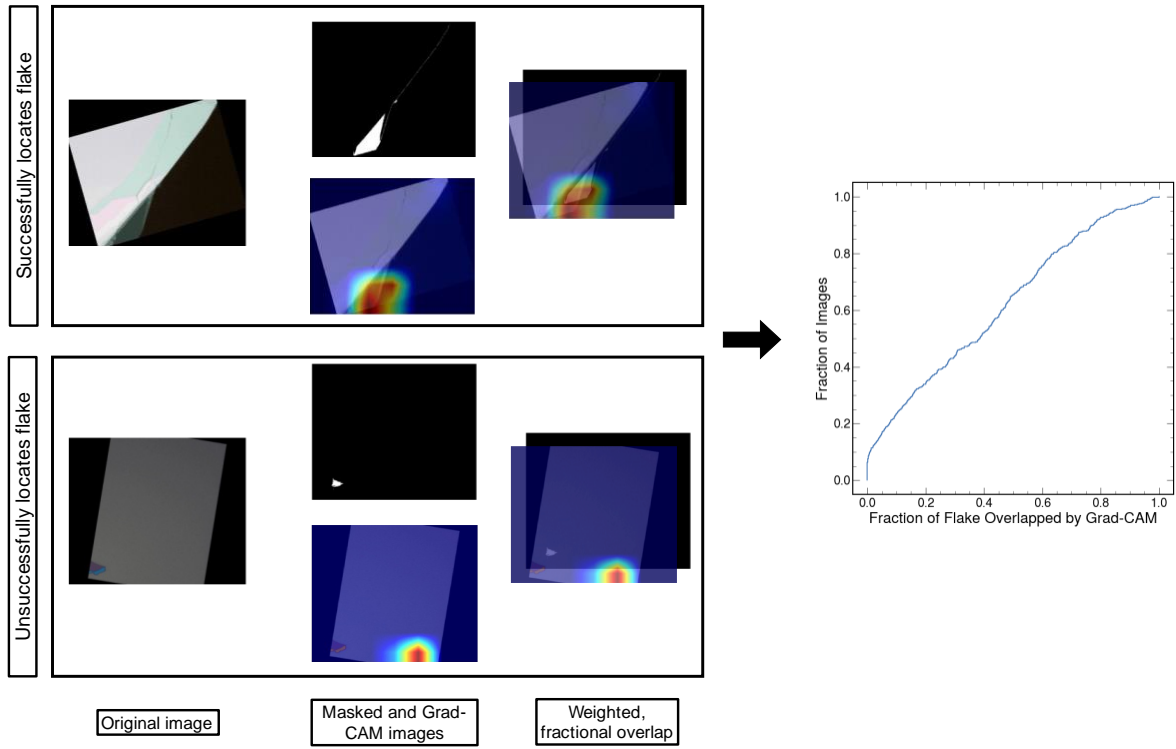

**Figure 3:** Grad-CAM Evaluation with an Empirical Cumulative Distribution Function. For 500 correctly classified flake images quantized with 20 colors, the weighted fraction of flake highlighted by the Grad-CAM heatmap is displayed as an empirical cumulative distribution function (on the right). The process for determining the fractional overlap for an unsuccessful image (overlap < 0.1) and a successful image (overlap > 0.9) are shown. In both instances, a masking algorithm locates the thin flake and determines the weighted fraction of overlap between this flake and the Grad-CAM heatmap.

a.

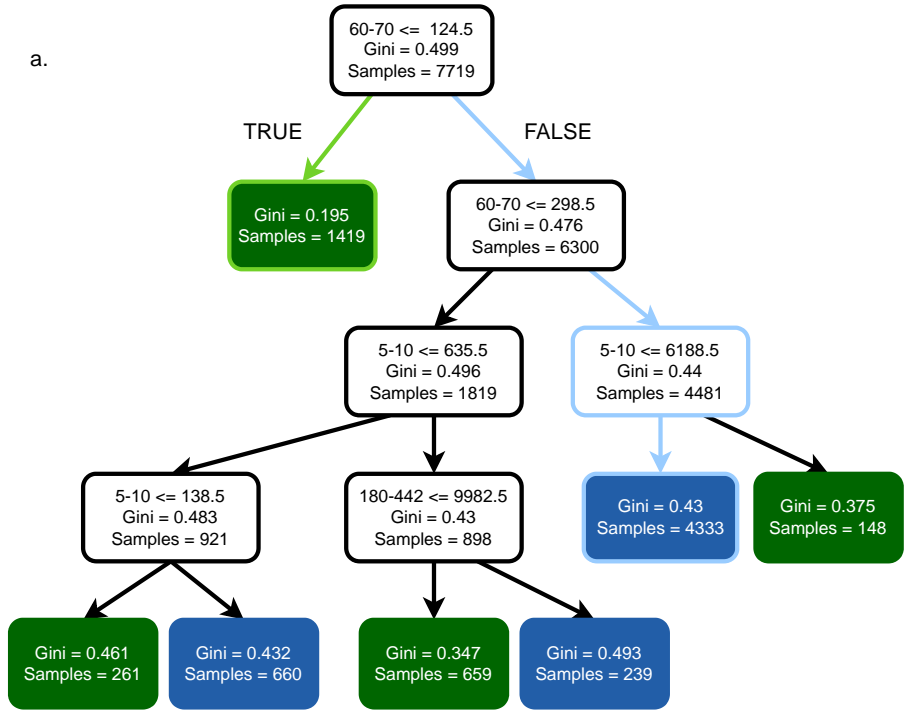

b.

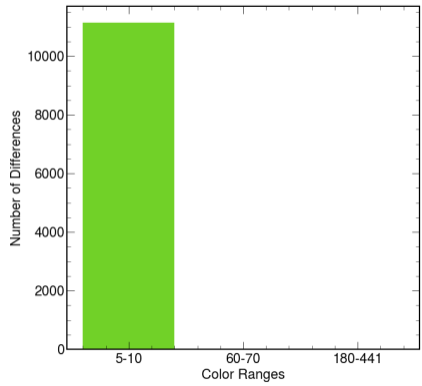

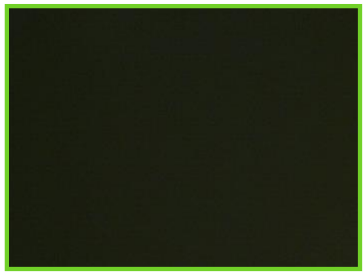

c.

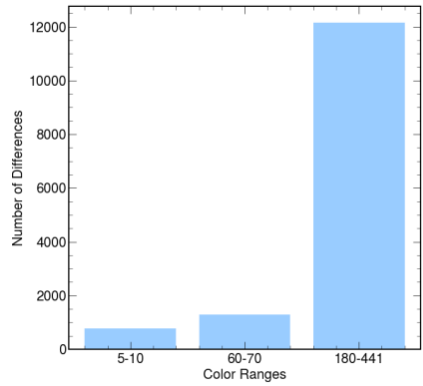

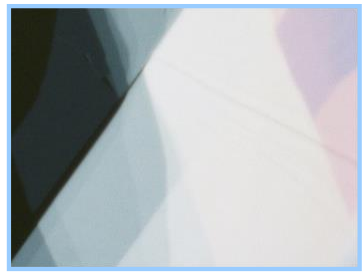

**Figure 4:** Visualization of decision tree after training. (**a**) Visualization of a decision tree classifier after training with 256 color clusters with a training and test accuracy of 73% and 67%. Each node shows the feature (number of color cluster differences in each range) used to make the binary decision, the Gini impurity (a measure of the probability of misclassifying a random element when randomly classifying it based on the class distribution), the node's samples, and classification if a terminal node (blue and green as flake and no flake). (**b**) An example of the tree traversal of an image without a flake and associated bins (colored light green) and (**c**) an image with a flake and associated bins (colored light blue).

## Data Augmentation

In this paper, four types of augmentation methods were applied sequentially to augment image data: padding, rotating, flipping, and color jitter. Augmented images have the same labels as the original images. The rationale of these augmentation methods is guaranteed by the belief that information from each image has some types of translation invariance. For instance, rotating or flipping an image does not change the existence of a flake or not. However, a dramatic modification of the brightness, contrast, or hue of an image (jitter) can remove flakes from images. To avoid mislabeling induced by data augmentation, we restricted the change of color jitter to be small.

We first padded each image by black bands on four sides, this step makes the image larger and thus helps avoid clipping flakes in the following rotation step. This rotation step randomly chooses a rotating degree uniformly from 0 to 360 degrees. After that, flip the image horizontally. Not all images are invariant to horizontal flipping, however, in our application it is a valid tool. In general, horizontal flipping is used much less than vertical. Finally, color, brightness, and hue of the image is changed by sampling corresponding changing factors uniformly from [0.9, 1.1]. For each image, we generated 30 independent augmented images, which enlarges data size by 30 times. When combining with the quantization method, we quantized the augmented images.

## Color Quantization

We employed color quantization using k-means from scikit-learn 1.1.1. The code utilized for this process was slightly modified by removing nested for loops and using torch.randperm from torch.tensor instead of sklearn.shuffle when fitting on a small section of the image.

## Tree-based Methods Features

Images were recreated with a selected number of k-means color centroids. All possible differences in these color centroids were placed into bins. The color difference between two colors was calculated as the square root of the squared difference between their RGB values. The following outlines this binning process. The lowest bin contained flake color differences that ranged from the lowest one calculated to 5. From there the color differences are binned in increments of 10 until an upper range of 140. A manual sampling of randomized images indicated that a flake to background differences would not exceed 140 therefore we selected a large bin for this upper range (140-180). The final bin encompasses all other color differences (140 – max color difference). The lower ranges effectively distinguished images that contained mainly background and thus have very similar colors creating consistently small differences. Color differences in middle ranges indicated the presence of flakes. Large color differences implied high amounts of bulk material. Specifically all calculated color ranges were: [minimum color difference - 5], [5 - 10], [10 - 20], [20 - 30], [30 - 40], [40 - 50], [50 - 60], [60 - 70], [70 - 80], [80 - 90], [90 - 100], [100 - 110], [110 - 120], [120 - 130], [130 - 140], [140 - 180], [180 - maximum color difference]. To prevent extreme overfitting of the tree-based classifiers when determining accuracy and confusion matrices, we only used the ranges [5,10], [60-70] and [180 – maximum color difference]. When creating receiver operating characteristic curves all ranges were used to better understand the estimators' sensitivity and specificity.

## Tree-based Methods Grid Search Cross-Validation

To determine the best hyperparameters for each tree-based method (decision tree, boosted decision tree, random forest) we employ the scikit learn GridSearchCV algorithm. This method performs an exhaustive search for optimization of an estimator's parameters by cross-validation over a parameter grid. For the decision tree estimator, the parameter grid consisted of the tree's max depth varying from one to 10 and its max leaf nodes varied from one to 10. For the boosted decision tree, we varied tested learning rates of 0.001, 0.01, and 0.1 and varied the tree's max depth from one to 10. For the random forest, the max depth varied from four to 10 and the number of n-estimators varied from one to 10.

## Masking Algorithm

We created a masking algorithm which when given an image with a thin flake will return an image with the flake in white and everything else in black. The algorithm accepts user input to select the flake in the image. We employed this masking algorithm for a post processing of images that the CNNs correctly classified.

## Confusion Matrices and Receiver Operating Characteristic Curves

The confusion matrices indicated that the tree-based algorithm successfully filtered out images without flakes while mainly misclassifying images with flakes as not containing them. Through manual identification, these false negatives usually contained very small pieces of flakes. However, in practice, a dataset of images of a slide containing a fabricated flake will contain a large majority of images without flakes rendering the algorithm's ability to filter out these images extremely useful.

Furthermore, the receiver operating characteristic (ROC) curves demonstrated that the tree-based algorithms can be tuned to have high true positive rates at the expense of moderately high false positive rates. This way the algorithm can accurately identify images with flakes and an experienced user can quickly go through these images to find the false positives. For example, in the ROC curve with 256 colors the random forest can achieve a true positive rate of about 80% and a false positive rate of about 20%. For example, if there are 100 images a realistic distribution would be 80 images with no flake and 20 with a flake. Once this dataset has been run through the tree-based algorithm, 16 images would be true positives and 16 images would be false positives. One would only have to go through 32 images (about a third of the original dataset) to find 80% of the flakes. In practice, this will rapidly speed up flake identification. Overall, both confusion matrices and ROC curves demonstrated that the ensemble algorithms performed with higher success than the single decision tree.

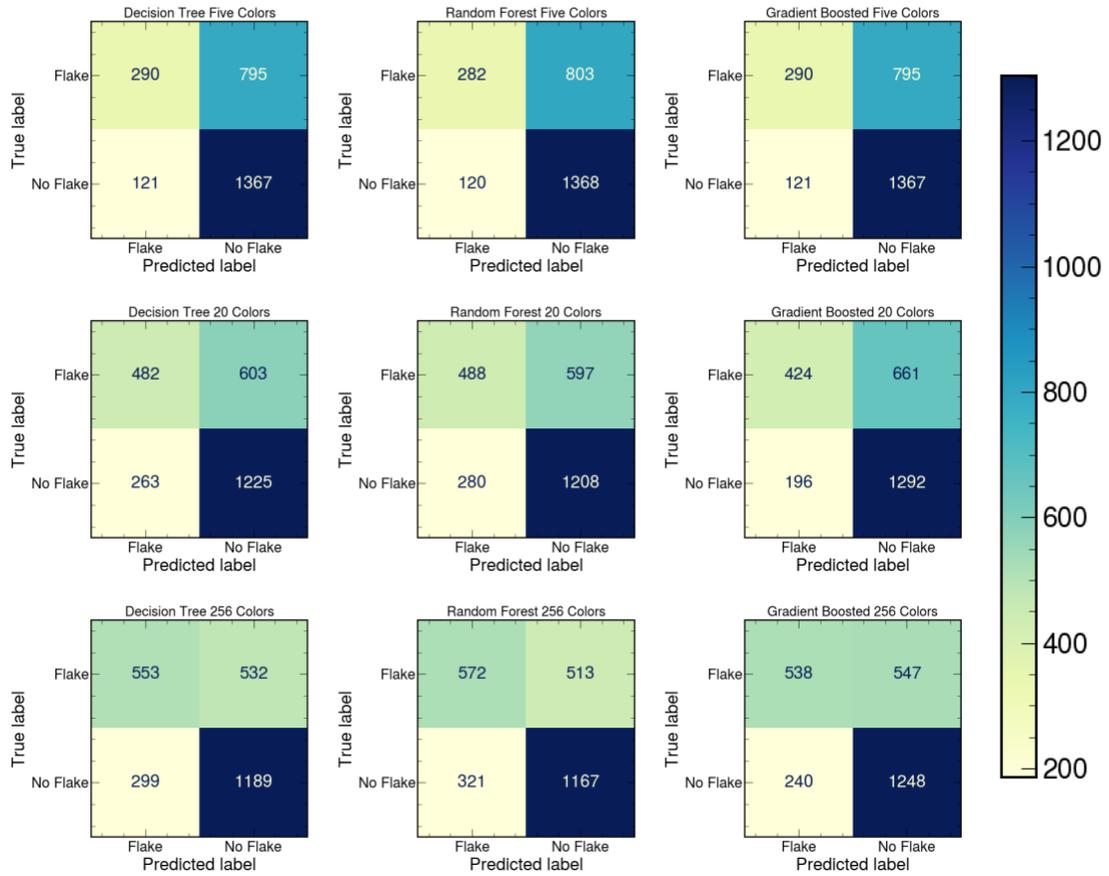

Figure 1: Confusion matrices for tree-based algorithms. Confusion matrices for tree-based classifiers when images are recreated with 5, 20, and 256 colors.

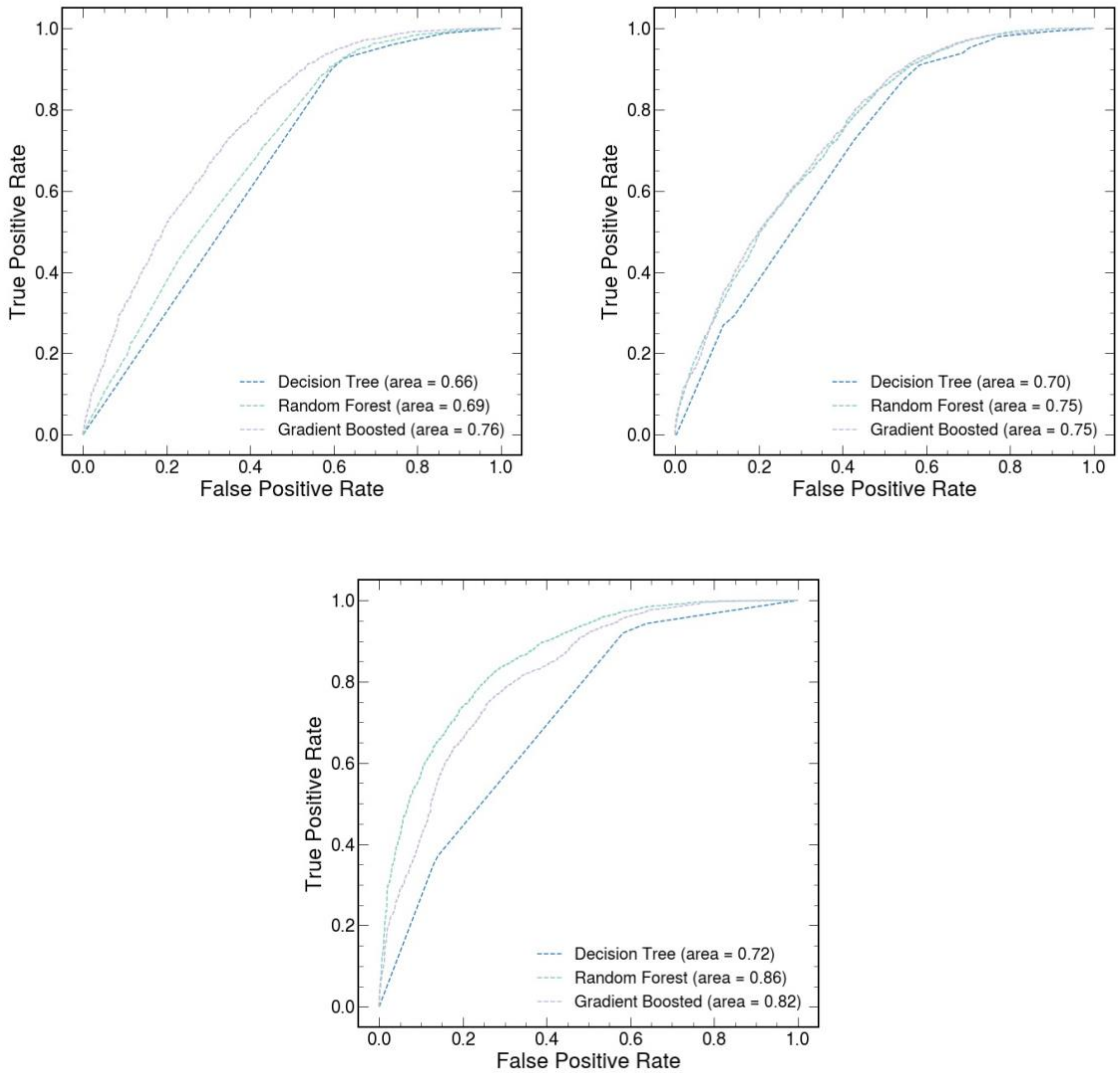

Figure 2: Receiver operating characteristic (ROC) curves. ROC curves for each tree-based classifier when images are recreated with 5, 20, and 256 colors.

## Classifier Accuracy on Subset Training

Deep-learning algorithms require large datasets while tree-based methods do not. To evaluate the effect of training dataset size on all algorithms, we trained the estimators with subsets of the training data after a standard 75/25 train test split. We trained on 10% and 50% of the available training data (75% of the original dataset). The accuracy of the CNNs was determined by their performance on the test data (25% of the original dataset). The tree-based methods were optimized with a grid search and 5-fold cross-validation. Once the hyperparameters were determined, the estimator's accuracy was determined from training and testing on available data. The CNNs accuracy decreased substantially with less training data while the tree-based methods maintained their performance.

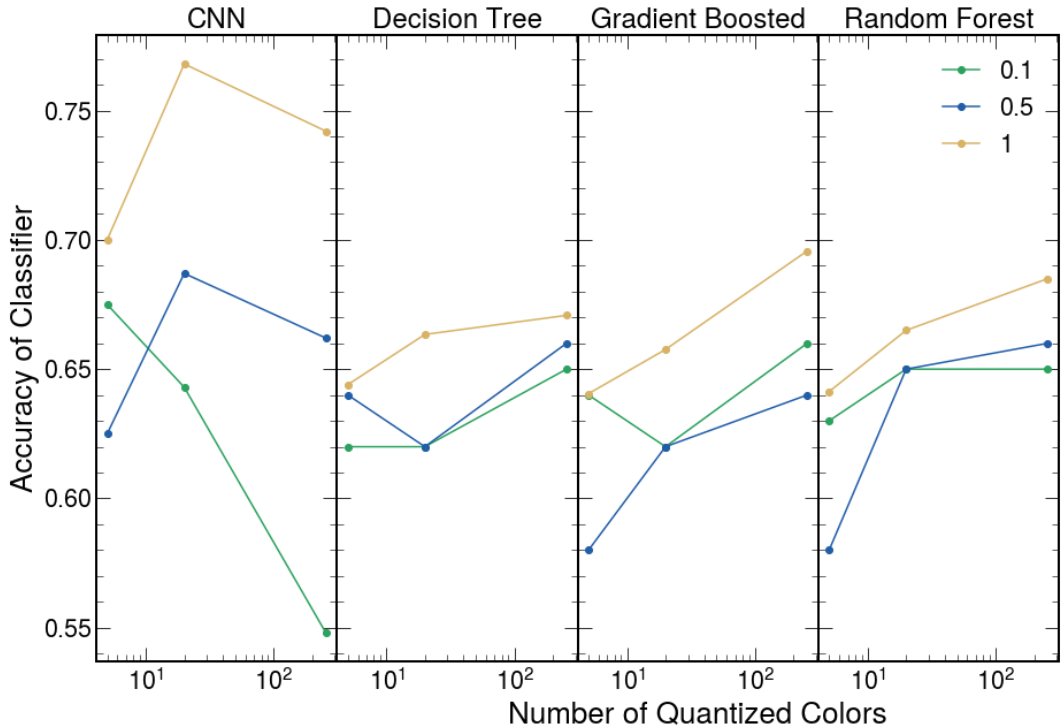

Figure 3: Accuracy of CNNs and tree-based methods when training on data subsets. CNNs' (first panel) and tree-based methods' (last three panels) test accuracies when trained with 10%, 50% and 100% of available training data from a 75/25 train test/split on entire data set.

**Table S1. Tree-based estimators' performance with 10% of training data and varied quantized colors**

| | | | | Quantized Colors | | | | | |
|---|---|---|---|---|---|---|---|---|---|
| | | 5 | | | 20 | | | 256 | |
| | Train | Test | Hyperparameters | Train | Test | Hyperparameters | Train | Test | Hyperparameters |
| Decision Tree | 0.66 | 0.62 | Max depth: 4 Max leaf: 4 | 0.79 | 0.62 | Max depth: 6 Max leaf: 4 | 0.87 | 0.65 | Max depth: 5 Max leaf: 7 |
| Gradient Boosted Decision Tree | 0.68 | 0.64 | Max depth: 2 Learning rate: 0.001 | 0.79 | 0.62 | Max depth: 3 Learning rate: 0.001 | 0.87 | 0.66 | Max depth: 3 Learning rate: 0.001 |
| Random Forest | 0.72 | 0.63 | Max depth: 4 N-estimators: 4 | 0.80 | 0.65 | Max depth: 4 N-estimators: 4 | 0.87 | 0.65 | Max depth: 4 N-estimators: 4 |

Fractional train and test accuracy and associated hyperparameters for tree-based estimators trained with 10% of available training data from a 75/25 train/test split.

**Table S2. Tree-based estimators' performance with 50% of training data and varied quantized colors**

| | Quantized Colors | | | | | | | | |
|---|---|---|---|---|---|---|---|---|---|
| | 5 | | | 20 | | | 256 | | |
| | Train | Test | Hyperparameters | Train | Test | Hyperparameters | Train | Test | Hyperparameters |
| Decision Tree | 0.61 | 0.64 | Max depth: 5 Max leaf: 4 | 0.72 | 0.62 | Max depth: 5 Max leaf: 4 | 0.74 | 0.66 | Max depth: 5 Max leaf: 4 |
| Gradient Boosted Decision Tree | 0.65 | 0.58 | Max depth: 7 Learning rate: 0.1 | 0.71 | 0.62 | Max depth: 3 Learning rate: 0.001 | 0.77 | 0.64 | Max depth: 4 Learning rate: 0.001 |
| Random Forest | 0.65 | 0.58 | Max depth: 10 N-estimators: 10 | 0.73 | 0.65 | Max depth: 5 N-estimators: 7 | 0.78 | 0.66 | Max depth: 5 N-estimators: 7 |

Fractional train and test accuracy and associated hyperparameters for tree-based estimators trained with 50% of available training data from a 75/25 train/test split.

**Table S3. Tree-based estimators' performance with all training data and varied quantized colors**

| | Quantized Colors | | | | | | | | |
|---|---|---|---|---|---|---|---|---|---|
| | 5 | | | 20 | | | 256 | | |
| | Train | Test | Hyperparameters | Train | Test | Hyperparameters | Train | Test | Hyperparameters |
| Decision Tree | 0.62 | 0.64 | Max depth: 5 Max leaf: 4 | 0.68 | 0.66 | Max depth: 7 Max leaf: 7 | 0.73 | 0.67 | Max depth: 7 Max leaf: 7 |
| Gradient Boosted Decision Tree | 0.61 | 0.64 | Max depth: 4 Learning rate: 0.001 | 0.67 | 0.66 | Max depth: 4 Learning rate: 0.001 | 0.74 | 0.70 | Max depth: 7 Learning rate: 0.001 |
| Random Forest | 0.62 | 0.64 | Max depth: 5 N-estimators: 7 | 0.68 | 0.67 | Max depth: 5 N-estimators: 7 | 0.73 | 0.69 | Max depth: 5 N-estimators: 7 |

Fractional train and test accuracy and associated hyperparameters for tree-based estimators trained with 100% of available training data from a 75/25 train/test split.

**Table S4. CNNs' performance with varied training data and quantized colors**

| | Training Data Used During Training | | | | | |
|---|---|---|---|---|---|---|
| | **10%** | | **50%** | | **100%** | |
| **Quantized Colors** | **Train** | **Test** | **Train** | **Test** | **Train** | **Test** |
| 5 | 1.0 | 0.68 | 1.0 | 0.63 | 1.0 | 0.7 |
| 20 | 1.0 | 0.64 | 1.0 | 0.69 | 1.0 | 0.77 |
| Infinite (all) | 0.96 | 0.54 | 1.0 | 0.67 | 1.0 | 0.74 |

Fractional train and test accuracy for CNNs' from 75/25 train/test split.